\documentclass[12pt]{iopart}

\begin{document}

\title{Primordial Intermediate Mass Black Holes\\ as Dark Matter}

\author{Paul H. Frampton}

\address{\it Dipartimento di Matematica e Fisica "Ennio De Giorgi",
Universit\`{a} del Salento and INFN-Lecce, Via Arnesano, 73100 Lecce, 
Italy.}
\ead{paul.h.frampton@gmail.com}
\vspace{10pt}
\begin{indented}
\item[]April 2020
\end{indented}

\begin{abstract}
Among particle theory candidates for the dark matter constituents. axions and WIMPs
are the most popular. In this talk we discuss these then focus on our preferred
astrophysical candidate, the Primordial Intermediate Mass Black Holes in the acronym
$DM=PIMBHs$. The earliest experimental confirmation may come from microlensing
of the Magellanic Clouds at the LSST 8m telescope in the mid-2020s, or possibly a few years earlier
in 2021 from work being pursued, using DECam data from the smaller Blanco 4m tescope,
at LLNL.
\end{abstract}

\large

\section{Introduction}

\noindent
Astronomical observations have led to a consensus that the energy
make-up of the visible universe is approximately 72\% dark energy, 24\% dark matter
and only 4\% normal matter. 

\bigskip

\noindent
General discussions of the history and experiments for dark matter are in
several books.
A popular book,
''The Cosmic Cocktail" by Katherine Freese, discusses a large number of
WIMP searches. 

\bigskip

\noindent
The present ignorance of the dark matter sector is put into perspective
by looking at the uncertainty in the values of the constituent
mass previously considered. The lightest such candidate is the ultra-light axion
with $M = 10^{-22} eV$. The heaviest such candidate for dark
matter, suitably defined in terms of the entropy of the universe, is
a supermassive mass
black hole (SMBH) with $M = 10^{12} M_{\odot}$. This is
one hundred orders of magnitude, $10^{100}$ or a googol, times more massive than the
ultra-light axion.

\bigskip

\noindent
Our aim
is to reduce this uncertainty.

\bigskip

\noindent
The result of the present analysis will be that the number of
orders of magnitude uncertainty in the dark matter constituent
mass can be reduced to two. We shall conclude,
after extensive discussion, that the most viable candidate for the constituent
which dominates dark matter in the Milky Way dark halo is \cite{C,F,CF}
the Primordial Intermediate Mass Black Hole (PIMBH) with mass in the range

\begin{equation}
25 M_{\odot} < M_{PIMBH} < 1,600 M_{\odot}
\label{IMBH}
\end{equation}

\noindent
corresponding to microlensing light curves of duration between one and eight
years  for the Blanco 4m telescope with DECam
at Cerro Tololo, Chile, pointed towards the Large Magellanic Cloud.

\bigskip

\noindent
An explanation for the relative neglect of PIMBHs, relative to WIMPs, may be that the literature is confusing.

\bigskip

\noindent
At least one study claimed to rule out the entire mass range
from $25M_{\odot}$ to $1600M_{\odot}$ that we displayed in Eq.(1). 
We shall attempt
to clarify the observational situation which actually still permits the whole range in Eq.(1).

\bigskip

\noindent
The present talk is, in part, an
attempt to redress the imbalance between the relatively few
experimental efforts to search for PIMBHs
compared to the very extensive variety of WIMP searches. 

\bigskip

\noindent
\section{Axions}

\bigskip
\bigskip

\noindent
It is worth reviewing briefly the history of the QCD axion particle now believed,
if it exists, to lie in the mass range

\begin{equation}
10^{-12} eV < M < 10^{-3} eV 
\label{axion}
\end{equation}

\noindent
The lagrangian originally proposed for Quantum Chromodymamics (QCD) was of the
simple form, analogous to Quantum Electrodynamics,

\begin{equation}
{\cal L}_{QCD} = -\frac{1}{4} G_{\mu\nu}^{\alpha}G^{\mu\nu}_{\alpha}
- \frac{1}{2}  \sum_i \bar{q}_{i,a}\gamma^{\mu}D^{ab}_{\mu} q_{i,b}
\label{QCD}
\end{equation}

summed over the six quark flavors.

\newpage

\noindent
The simplicity of Eq.(\ref{QCD}) was only temporary and became more
complicated
in 1975 by the discovery of instantons
which dictated an additional term 

\begin{equation}
\Delta {\cal L}_{QCD} = \frac{\Theta}{64\pi^2} G_{\mu\nu}^{\alpha} \tilde{G}^{\mu\nu}_{\alpha}
\label{GGdual}
\end{equation}

\noindent
where $\tilde{G}_{\mu\nu}$ is the dual of $G_{\mu\nu}$. 
The additional term, Eq.(\ref{GGdual}),violates CP and contributes to the neutron
electric dipole moment whose upper limit  provides a constraint on $\bar{\Theta}$
(simply related to $\Theta$)

\begin{equation}
\bar{\Theta} < 10^{-9}
\label{strongCP}
\end{equation}

\noindent
The QCD axion arises from spontaneously breaking a symmetry
imposed to set $\bar{\Theta}=0$. The QCD axion
may exist and contribute to dark matter.

\bigskip

\noindent
Recently there has been widespread interest in axion-like particles(ALPs).
For dark matter, there is the possibility of ultra-light ALPs with masses as low as
$10^{-22} eV$. 

\bigskip

\noindent
Such ultra-light axions are suggested by moduli and dilatons in string theory
and can ameliorate some issues for cold dark matter(CDM), cusps 
at galactic centres and excess satellite galaxies associated with CDM calculaltions.

\bigskip

\noindent
These ultra-light axions play no role in solving the strong CP problem and so "axion"
is an slightly inappropriate name; ulta-light boson is better. Such ultra-light bosons can superradiate
and limit the spin of Kerr black holes.

\bigskip

\section{WIMPs}

\bigskip

\noindent
By Weakly Interacting Massive Particle (WIMP) is generally meant an unidentified elementary
particle with mass in the range, say, between 10 GeV and 1000 GeV and with scattering cross
section with nucleons ($N$) satisfying, according to the latest unsuccessful WIMP direct searches, 

\begin{equation}
\sigma_{WIMP-N} < 10^{-45} cm^2
\label{WIMPcc}
\end{equation}

\bigskip
\bigskip

\noindent
which is roughly comparable to the characteristic strength of the known weak interaction.
Actually, Eq.(\ref{WIMPcc}) now imposed by WIMP searches is a few orders
of magnitude below the expectation from weak interactions. This may be a first
sign that something is rotten in the state of Denmark.

\bigskip

\noindent
The WIMP particle must be electrically neutral and be stable or have an extremely
long lifetime. In model-building, the stability may be achieved by an {\it ad hoc} discrete
symmetry, for example a $Z_2$ symmetry under which all the standard model
particles are even and others are odd. If the discrete symmetry is unbroken,
the lightest odd state must be stable
and therefore a candidate for a dark matter. In general, this appears contrived because
the discrete symmetry is not otherwise motivated.

\bigskip

\noindent
In sypersymmetry, such a discrete symmetry appears
naturally and gives rise to a wonderful candidate for a WIMP called the neutralino.
However, since the LHC data has lent no support to weak-scale supersymmetry,
this WIMP has less motivation than it once did.

\bigskip

\section{MACHOs}

\bigskip

\noindent
Massive Compact Halo Objects (MACHOs) are commonly defined
by the notion of compact objects used in astrophysics as
the end products of stellar evolution when most of the nuclear fuel has been expended.
They are usually defined to include white dwarfs, neutron stars, black holes, brown dwarfs
and unassociated planets, all equally hard to detect because they do not emit any
radiation.

\bigskip

\noindent
This narrow definition implies, however, that MACHOs are composed of normal matter
which is too restrictive in the case of black holes.
It has been shown that black holes of
mass up to $100,000 M_{\odot}$ (even up to $10^{12}M_{\odot}$) can be
produced primordially, according to a paper published at IPMU-Univ. of Tokyo
by F.K.T.Y. in 2010\cite{FKTY,FK}.

\bigskip

\noindent
Unlike the axion and WIMP elementary particles which would have a definite mass, the
black holes will have a range of masses. The lightest PBH which has
survived for the age of the universe has a lower mass limit

\begin{equation}
M_{PBH} > 10^{-18} M_{\odot} \sim 10^{36} TeV
\label{PBHmin}
\end{equation}
\noindent
already thirty-six orders of magnitude heavier than the heaviest would-be WIMP.
This lower limit comes from the lifetime formula derivable from Hawking radiation

\begin{equation}
\tau_{BH}(M_{BH}) \sim \frac{G^2 M_{BH}^3}{\hbar c^4} 
\sim 10^{64} \left( \frac{M_{BH}}{M_{\odot}} \right)^3  years
\label{BHlfetime}
\end{equation}

\bigskip

\noindent
Because of observational constraints 
the dark matter
constituents must generally be another twenty orders of magnitude more massive
than the lower limit in Eq.(\ref{PBHmin}).  

\bigskip

\noindent
We assert that most dark
matter black holes are in the mass range above
25 and up to 1,600 and more times the solar mass.
The name primordial intermediate mass black holes (PIMBHs)
is appropriate because they lie in mass above stellar-mass black holes                                                                                                                                                                                             and below the
supermassive black holes which reside in galactic cores.

\bigskip

\noindent
Let us discuss three methods (there may be more)
which could
be used to search for dark matter PIMBHs. 
While so doing we shall clarify 
what limits, if any, can be deduced from
present observational knowledge.

\bigskip

\noindent
Before proceeding, it is appropriate first
to mention the important Xu-Ostriker upper bound 
of $10^5 M_{\odot}$ from galactic disk stability
for any MACHO residing inside the Milky Way galaxy.

\subsection{Wide Binaries}

\bigskip

\noindent
There exist in the Milky Way pairs of stars which are gravitationally bound binaries
with a separation more than 0.1pc. These wide binaries retain their original orbital parameters
unless compelled to change them by gravitational influences, for example, due to
nearby IMBHs.

\bigskip

\noindent
Because of their very low binding energy, wide binaries are particularly sensitive 
to gravitational perturbations and can be used
to place an upper limit on, or to detect, IMBHs. 
The history of employing this ingenious technique is regretfully checkered. 
In 2004 a fatally strong constraint was claimed by an Ohio State University group
in a paper entitled
"End of the MACHO Era" so that, for researchers who have time
to read only titles and abstracts, stellar and higher mass constituents of
dark matter appeared to be totally excluded.

\bigskip

\noindent
Five years later in 2009, however, another group this time from Cambridge University
reanalyzed the available data on wide binaries
and reached a quite different conclusion.
They questioned whether {\it any} rigorous constraint on MACHOs
could yet be claimed, especially as one of the important binaries
in the earlier sample had been misidentified.

\bigskip

\noindent
Because of this checkered history, it seems wisest to proceed with
caution but to recognize that wide binaries represent a potentially useful
source both of constraints on, and the possible discovery of, 
dark matter IMBHs.

\subsection{Distortion of the CMB}

\bigskip

\noindent
This approach hinges on the phenomenon of accretion of gas onto the PIMBHs.
The X-rays emitted by such accretion of gas are downgraded in frequency
by cosmic expansion and by Thomson scattering becoming microwaves which 
distort the CMB, both with regard to its spectrum and to its anisotropy.

\bigskip

\noindent
One impressive-seeming calculation by Ricotti, Ostriker and Mack (ROM) in 2008
of this effect employed a specific model for the
accretion, the Bondi model, and carried through the computation
all the way up to a point of comparison with data from FIRAS on CMB spectral distortions,
where FIRAS was a sensitive device attached to the COBE satellite.

\bigskip

\noindent
Unfortunately the Bondi model was invented for a static object
and assumes spherically symmetric purely s-wave accretion with radial inflow. Studies 
of the SMBH in the giant galaxy M87 have shown since 2014 that 
higher angular momenta strongly dominate, not surprising as the
SMBH possesses a gigantic spin angular momentum in natural units.

\bigskip
\bigskip

\noindent
The results from M87 suggest the upper limits
on MACHOs imposed by ROM
were too severe by orders of magnitude
and that up to 100\% of the dark matter is
permitted o be in the form of PIMBHs. In 2016, Ostriker privately
withdrew the ROM limit as being "far too severe". A more recent 2017
modified version of this calculation (Ali-Ha\"{i}moud and Kamionkowski),
apparently unaware of ROM's withdrawal,
similarly overestimates the accretion by assuming quasi-sphericity
and arrives at far too strong bounds on PIMBHs.

\bigskip

\subsection{Microlensing}

\bigskip

\noindent
Microlensing is the most direct experimental method and has the big advantage
that it has successfully found
examples of MACHOs. The MACHO Collaboration used a method
which had been proposed\footnote{We have read that such gravitational lensing was later found to have been calculated 
in unpublished 1912 notes by Einstein who did not publish perhaps because at that time he considered its experimental measurement
impracticable.} by Paczynski where the amplification of a distant
source by an intermediate gravitational lens is observed. The MACHO Collaboration
discovered several striking microlensing events whose light curves are
exhibited in its 2000 paper. The method certainly worked well for $M <  10 M_{\odot}$
and so should work equally well for $M > 25 M_{\odot}$.

\bigskip

\noindent
The longevity of a given lensing event is proportional to the square root of the lensing mass
and, in an admittedly crude approximation, is given by
($\hat{t}$ is duration)

\begin{equation}
\hat{t} \simeq 0.2 yr \left( \frac{M_{lens}}{M_{\odot}} \right)^{1/2}
\label{that}
\end{equation}
\noindent
where a transit velocity $200km/s$ is assumed for the lensing object.

\bigskip

\noindent
The MACHO Collaboration investigated lensing events with durations
ranging between about two hours and 200 days. From Eq.(\ref{that}) this corresponds
to MACHO masses between approximately $10^{-6} M_{\odot}$ and $10 M_{\odot}$.

\bigskip

\noindent
The total number and masses of objects discovered by the MACHO Collaboration
could not account for all the dark matter known to exist in the Milky Way. At most
10\% could be explained. To our knowledge, the experiment ran out of money and
was essentially abandoned in about the year 2000.
But perhaps the MACHO Collaboration and its funding
agency were too easily discouraged.

\bigskip

\noindent
What is being suggested is that the other 90\% of the dark matter in the
Milky Way is in the form of MACHOs which are more massive than those detected
by the MACHO Collaboration, and which almost certainly could be detected by a
straightforward extension of their techniques. In particular, the expected
microlensing events have
a duration ranging from one to eight years.

\bigskip

\noindent
Microlensing experiments involve systematic scans of millions of distant star
sources because it requires accurate alignment of the star and the
intermediate lensing MACHO. The experiments are highly computer
intensive, and requires a sophisticated data pipeline.

\bigskip

\noindent
The experiment is undoubtedly extremely
challenging, but there seems no obvious reason it is impracticable.
Certain new hurdles have already been discovered ({\it e.g.} ''crowding")
 but there is reason to think that by 2021
a definitive paper might be published by the LLNL group.

\bigskip

\subsection{Interim discussion}

\bigskip

\noindent
Axions may not exist for theoretical reasons
discovered in 1992. Electroweak supersymmetry probably does not exist
for the experimental reason of its non-discovery 
at the LHC.
The idea that dark matter experiences weak interactions (WIMPs) came historically
from the appearance of an appealing DM constituent, the neutralino,
in the theory of electroweak supersymmetry for which there is no experimental evidence.

\bigskip

\noindent
The only interaction which we know for certain to be experienced
by dark matter is gravity and the simplest assumption is that gravity
is the only force coupled to dark matter.
Why should the dark matter experience the weak interaction
when it does not experience the strong and electromagnetic interactions?

\bigskip

\noindent
All terrestrial experiments searching for dark matter by either direct
detection or production may be doomed to failure. 

\bigskip

\noindent
We began with four candidates for dark matter constituent:
(1) axions; (2) WIMPs; (3) baryonic MACHOs;
(4) PIMBHs.
We disfavoured the first two by
arguments made within the context 
of particle phenomenology. We eliminated the third by the upper limit
on baryons imposed by robust BBN calculations.

\bigskip

\noindent
Contrary to claims made in the dark matter literature $f_{DM}=1$ is not excluded 
for $25M_{\odot} < M_{PIMBH} < 10^5 M_{\odot}$.
Exclusion plots in the literature which disagree with this assertion make 
unreliable assumptions on accretion, such as a Bondi model
with spherical symmetry and radial inflow.

\bigskip

\noindent
We assert that PIMBHs can constitute almost all  dark matter while maintaining
consistency with the BBN calculations. 

\bigskip

\noindent
Our proposal is that the Milky Way contains between ten million and ten billion
massive black holes each with between a hundred and a hundred thousand times the solar mass. Assuming the halo
is a sphere of radius a hundred thousand light years the typical separation
is between one hundred and one thousand light years which is also the most
probable distance of the nearest PIMBH to the Earth. At first sight, it may be
surprising that such a huge number of PIMBHs
-- the plums in a {\it ``PIMBH plum pudding"} --(c.f. Thomson 1904) could remain undetected.
2015 was 111 years after Thomson and the halo 31 powers of ten bigger
than the atom.
However, the mean
separation of the plums
is at least a hundred light years and the plum size
is smaller than the Sun.

\bigskip

\noindent
Of the detection methods discussed, extended microlensing observations
seem the most promising and an experiment to detect
higher longevity microlensing events is being actively pursued.
The wide-field telescope must be in the Southern Hemisphere
to use the Magellanic Clouds (LMC and SMC) for sources.

\section{Microlensing Experiments}

\bigskip

\noindent
The most appropriate telescope for studying the MCs, active since 1986, is
the Blanco 4m at Cerro Tololo with its DECam
having 520 Megapixels.
This telescope was named after the late Victor Blanco,
the Puerto Rican astronomer who was the CTIO Director.
A bigger and more powerful telescope
will be the LSST (= Large Synoptic
Survey Telescope) under construction in Northern Chile which
will take first light in 2022. It will be 8.4m with a 3200 Megapixel
camera, clearly superior to the Blanco 4m. Since, however, we require 
two-year duration light curves
to discover $100M_{\odot}$ lenses, the earliest that
LSST could discover convincing evidence for DM=PIMBHs
would be in 2024 or later.

\bigskip

\noindent
An ongoing MC microlensing project which includes George Chapline (theorist) 
and eight experimentalists (which include the group's PI Will Dawson) is based
at LLNL.
Their data taking started over two years ago in February 2018. Their interesting
work has revealed a technical issue which could have been, but was not, anticipated.
The issue is stellar merging which arises, relative to the celebrated experiment
done in the 20th century by Alcock, {\it et al.} (The MACHO Collaboration)\cite{Alcock}, for two main reasons. 
We recall that Alcock,{\it et al.} looked at
MACHOs only with mass below $20M_{\odot}$ while the LSST will look for lenses with
significantly higher masses.
First, the Einstein radius increases with the lens mass $\propto M^{\frac{1}{2}}$;
second, the longer exposure as the light curve duration increases (also 
$\propto M^{\frac{1}{2}}$) probes more deeply and records lower magnitude stars. 
Both effects increase the number of stars and hence exacerbate stellar merging.

\bigskip

\noindent
Resolution of the issue should be possible using the techniques of CFP
(=Crowded Field Photometry), a mature area of research which goes back
over thirty years\cite{Stetson}. The experimental situation may thus be summarised
as the expectation that the LSST will settle this speculation about dark
matter by the mid-2020s but they could be scooped by years
if the LLNL group can successfully employ CFP to resolve stellar merging,
conceivably as early as 2021.

\section{The Reason Dark Matter Exists}

\noindent
The papers cited in the Introduction \cite{C,F,CF} address
the question of what are the constituents of Dark Matter, and the
suggested answer to that question is they are our titular Primordial
Intermediate Mass Black Holes.

\bigskip

\noindent
The question why dark matter exists at all seems to us to be equally as important
as what the dark matter is, so we have given a more complete
discussion about the origin and nature of the dark matter in \cite{F2}.
The answer lies the second law of thermodynamics applied to the entropy of the universe
during the time when the PIMBHs are formed.
There is no comparable reason for the formation of WIMPs or axions.

\bigskip

\noindent
The entropy from the SMBHs known to exist at galactic cores gives a contribution to the
dimensionless entropy $S/k$ of the universe which is roughly $S/k \sim 10^{103}$ or one thousand googols.
The identification $DM=PIMBHs$ can increase this already large number by as 
much as another factor of a thousand to $S/k \sim 10^{106}$, or one million googols,
depending on the PIMBH mass function.
The second law of thermodynamics applied to the cosmic entropy in the early 
universe is, we believe, why dark matter was originally formed, and still exists at present, in the form
of Primordial Intermediate Mass Black Holes.

\bigskip

\noindent
We eagerly await the verdict of Nature on our speculation, expected to be revealed at the LSST in the
mid-2020s, and conceivably much earlier by the important and interesting work presently being done at
Lawrence Livermore National Laboratory, in California.

\end{document}